\documentstyle[aps]{revtex}
\begin{document}
\draft
\title{One-dimensional classical adjoint SU(2) Coulomb Gas}
\author{Michael Engelhardt\thanks{email: 
engelm@pthp1.tphys.physik.uni-tuebingen.de}\thanks{Supported by
Deutsche Forschungsgemeinschaft under DFG Re 856 / 1-3} }
\address{Institut f\"ur theoretische Physik \\
Universit\"at T\"ubingen \\
Auf der Morgenstelle 14 \\ 72076 T\"ubingen, Germany }
\date{}
\maketitle

\begin{abstract}
The equation of state of a one-dimensional classical
nonrelativistic Coulomb gas of particles in the adjoint representation 
of SU(2) is given. The problem is solved both with and without
sources in the fundamental representation at either end of the
system. The gas exhibits confining properties at low densities
and temperatures and deconfinement in the limit of high densities and
temperatures. However, there is no phase transition to a regime
where the string tension vanishes identically; true deconfinement
only happens for infinite densities and temperatures. In the low
density, low temperature limit, a new type of collective behavior
is observed.
\end{abstract}
\vspace{1cm}


\section{Introduction}
Recently, there has been renewed interest in nonabelian classical
Coulomb gases in one space dimension \cite{zar1}-\cite{zar3}, in
particular ones which contain particles in the adjoint representation
of the gauge group. This is motivated by dimensional reduction
arguments which map the gluonic fields of QCD in two space dimensions
onto adjoint scalar fields in one space dimension \cite{kdb1}. Such
models are thus thought to contain some of the nontrivial gluonic
dynamics, e.g. the degrees of freedom of the color flux string,
of higher-dimensional Yang-Mills theories. While adjoint QCD
is not exactly solvable even in one space dimension, some of these
features have indeed been observed in approximate calculations
exploiting the limit of a large number of colors N$_{C} $ 
\cite{kdb1}-\cite{zhi}. If one further specializes to the limit of
heavy matter fields while retaining the large-N$_{C} $ limit, it
is possible to solve for the thermodynamic properties of the
one-dimensional gas of adjoint particles \cite{zar1}-\cite{zar3};
one finds a deconfinement phase transition as a function of
density or temperature \cite{zar1}-\cite{zar3},\cite{kuta}.

The purpose of the present note is to supplement this information with
the thermodynamics of the classical nonrelativistic adjoint Coulomb
gas in the opposite limit of only two colors. The thermodynamics
of analogous systems containing abelian particles \cite{len} or
particles in the fundamental representation of SU(2)
\cite{bam},\cite{me} has been known for some time and it turns out that
the case of adjoint particles can be solved with the same techniques as
employed in \cite{len},\cite{me}. Below, a pure adjoint Coulomb gas
as well as an adjoint Coulomb gas in the presence of two charges in
the fundamental representation of SU(2) are investigated. As is to be
expected, the system exhibits confining properties at low densities
and temperatures and deconfinement in the limit of high densities or
temperatures. In contradistinction to the large-N$_{C} $ case, the
two regions are not separated by a phase transition; there is a 
continuous crossover between the two types of behavior. This is
natural for a one-dimensional model with only a finite number of
degrees of freedom at each space point. The most interesting feature
is a new form of collective behavior at low densities and temperatures
which is not observed for particles in the fundamental representation
or abelian particles.

\section{The model}
In one space dimension, an array of color sources is connected by color
strings which begin and end at sources and may overlap (note that there
are no propagating gluon degrees of freedom in one space dimension).
If one thinks 
of the array of sources as being built up one by one from one end to the
other, then $n$ strings emanating from the first $k$ sources corresponds
to these sources being coupled to color spin $n/2$ in the case of SU(2)
color. Adding the $(k+1)$-th color source may change the number of color
strings according to the possible couplings of the $(k+1)$-th source to
the first $k$. E.g. in the case of SU(2) fundamental sources
\cite{bam},\cite{me}, $n\pm 1$ strings may emanate from the $(k+1)$-th
source; in the case of SU(2) adjoint sources, $n$ or $n\pm 2$ strings
may emanate from the $(k+1)$-th source since the adjoint sources carry
color spin one. In a physical system, the complete array of particles
acting as sources must be coupled to a color singlet, which also implies
that every different coupling scheme, i.e. string configuration, enters
the partition function with unit combinatorial weight.

The potential energy resulting from the formation of the strings between
the sources can be read off by observing that $n$ overlapping strings
are associated with an energy density of $\epsilon_{n} = g^2 n(n+2)/8$
(cf. e.g. \cite{berai}). Therefore, given the positions $q_i $ of $N$
particles and the numbers $n_i^C $ of strings between the $i$-th and the
$(i+1)$-th particle in a particular string configuration $C$, the associated
potential energy is
\begin{equation}
V_C = \frac{g^2 }{8} \sum_{i=1}^{N} n_i^C (n_i^C +2) (q_{i+1} -q_i )
\end{equation}
where $n_N^C=0$. Also defining $n_0^C=0$, one can write this as
\begin{equation}
V_C = \frac{g^2 }{8} \sum_{i=1}^{N} q_i ((n_{i-1}^{C} +1)^2 -(n_i^C +1)^2 )
\end{equation}
For a system consisting purely of particles in the adjoint representation,
the $n_i^C $ are subject to the constraints
\begin{equation}
n_0^C = n_N^C =0 \ \ \ \ \ \ n_i^C \ge 0 \ \ \ \ \ \
| n_{i+1}^{C} - n_i^C | \in \{ 0,\pm 2 \}
\end{equation}
The set of possible configurations $C$ is given by all sets 
$n_i^C , i=0,\ldots ,N $, which satisfy the above constraints.

\section{Thermodynamics}
The classical thermodynamics of a Coulomb gas of $N$ nonrelativistic adjoint
particles\footnote{The realm of validity of such a treatment is discussed
in \cite{me}.} is most easily calculated in the canonical constant
pressure ensemble \cite{len} (the physical reason for this is discussed
in \cite{prc}). In this ensemble, the Gibbs free energy after the trivial 
integrations over the canonical momenta is given by
\begin{equation}
G = -\frac{1}{\beta } \ln \left[ \left( \frac{2m\pi }{\beta } \right)^{N/2}
Q \right]
\end{equation}
with the configuration integral
\begin{eqnarray}
Q &=& \sum_{C} \int_{0}^{\infty } dq_{N} \int_{0}^{q_{N} } dq_{N-1}
\ldots \int_{0}^{q_2 } dq_1 \, e^{-\beta (Pq_{N} + V_C ) } \label{confi} \\
&=& \sum_{C} Q_{C}
\end{eqnarray}
where $P$ denotes the pressure. Due to the linear dependence of $V_C $
on the $q_i $, the integrations in (\ref{confi}) can all be carried
out using the properties of the Laplace transform \cite{prc}, yielding
\begin{equation}
Q_C = \left( \frac{8}{\beta g^2 } \right)^{N} (\gamma +1)
\frac{1}{\prod_{i=0}^{N} (\gamma + (n_i^C +1)^2 ) }
\label{produ}
\end{equation}
where the abbreviation $P=g^2 (\gamma +1)/8 $ was introduced. It remains
to evaluate the sum over configurations $C$. In order to accomplish this,
one defines a {\em cluster at level r } as a sequence of factors in
(\ref{produ}) such that $n_i^C \ge r $ for all $n_i^C $ in the
sequence, but $n_i^C < r $ for the two factors adjacent on either side.
This in particular implies $n_i^C =r $ for the first and the last
factor in the sequence; the value $r$ is not excluded for other
$n_i^C $ in between. Such a cluster thus physically corresponds
to a sequence of interparticle spacings covered by $r$ or more
strings\footnote{Note that the factors containing $n_0^C $ and
$n_N^C $ have been added by hand to the original $N-1$ physical
spacings.}. Then one can
define $G_n^r $ to be the sum over all clusters at level $r$
involving exactly $n$ factors. Ultimately, one is interested in
$G_{N+1}^{0} = (\beta g^2 /8)^{N} Q/(\gamma +1) $.

These cluster sums are convenient in that one can give a
recursion relation between the sums at different levels.
Excluding for the moment the case of level $r=0$, one can characterize
a cluster at level $r$ of length $n$ by the number $k$ of clusters at
level $r+2$ it contains (this is at most the integer part of $(n-1)/2$),
and the number $m$ of factors $(\gamma + (r+1)^2 ) $ present, i.e. the
number of spacings in the cluster covered by exactly $r$ strings.
This latter number is at least $k+1$ and at most $n-k$. Now, given
$k$ and $m$, one must have at least one factor $(\gamma + (r+1)^2 ) $
between each pair of clusters at level $r+2$ present, and one at
each end of the cluster at level $r$ under consideration. Thus,
$k+1$ of the $m$ factors $(\gamma + (r+1)^2 ) $ are fixed, but the
other $m-k-1$ factors can be arbitrarily distributed among the
$k+1$ spaces between the clusters at level $r+2$ or at the two ends.
The number of ways of doing this is $(m-1)!/(k! (m-1-k)!)$.
Finally, one must specify the lengths of the $k$ clusters at level
$r+2$ subject to the constraint that the sum of these lengths must
total $n-m$. Therefore, one has the recursion relation
\begin{equation}
G_n^r = \frac{1}{(\gamma + (r+1)^2 )^n } +
\sum_{k=1}^{[(n-1)/2]} \ \sum_{m=k+1}^{n-k} { m-1 \choose k }
\frac{1}{(\gamma + (r+1)^2 )^m } \
\sum_{i_1 =1}^{\infty } \ldots \sum_{i_k =1}^{\infty }
\delta_{i_1 + \ldots +i_k , n-m}
G_{i_1 }^{r+2} \ldots G_{i_k }^{r+2}
\label{recur}
\end{equation}
where the $k=0$ term has been written explicitly.
This recursion relation can be simplified by going to the
corresponding generating function
\begin{equation}
G^r (z) = \sum_{n=1}^{\infty } G_n^r z^n
\end{equation}
Treating the $n=1$ and $n=2$ terms separately
and rearranging summations,
\begin{equation}
\sum_{n=3}^{\infty } \sum_{k=1}^{[(n-1)/2]} \sum_{m=k+1}^{n-k} = 
\sum_{k=1}^{\infty } \sum_{m=k+1}^{\infty } \sum_{n=m+k}^{\infty }
\end{equation}
the Kronecker-$\delta $ constraint in (\ref{recur}) becomes trivial
and one arrives at
\begin{equation}
G^r (z) = \frac{1}{(\gamma +(r+1)^2 )/z -1 -G^{r+2} (z) }
\label{regf}
\end{equation}
As mentioned above, the case $r=0$ must be treated separately. One
cannot have consecutive factors $(\gamma +1)$ in a cluster at level
zero, i.e. consecutive interparticle spacings with no strings,
since color spin zero cannot be coupled with color spin one back
to color spin zero. Physically, this means that the adjoint particles
are confined\footnote{Note that this is not the case anymore in a
higher number of space dimensions, where a single adjoint particle can
form a color singlet bound state with a gluon.}.
Therefore, one has the constraint
\begin{equation}
m=k+1
\label{ct}
\end{equation}
and no $k=0$ term (one may set $G_1^0 = G_2^0 =0$; these quantities are
anyway irrelevant). Proceeding in complete analogy to above, one arrives
at
\begin{equation}
G^0 (z) = \frac{1}{(\gamma +1)/z -G^2 (z) } - \frac{z}{\gamma +1}
\label{inire}
\end{equation}
Now, one can reextract the $(N+1)$-th Taylor coefficient $G^0_{N+1} $
using
\begin{equation}
G^0_{N+1} = \frac{1}{2\pi i} \oint dz \, \frac{G^0 (z) }{z^{N+2} }
\end{equation}
By blowing up the integration contour to infinity, one picks up 
contributions from all the poles $\alpha_{i} $ of $G^0 (z) $,
\begin{equation}
G^0_{N+1} = -\sum_{i} \mbox{res}_{\alpha_{i} } \frac{G^0 (z) }{z^{N+2} }
\end{equation}
but for $N\rightarrow \infty $, this sum is dominated by the pole
closest to the origin\footnote{Note that the residue there is negative, 
since $G^0 (z) $ rises monotonously
on the real axis as it approaches $\alpha_{1} $.} $\alpha_{1} $,
\begin{equation}
\lim_{N\rightarrow \infty } \ln G^0_{N+1} =
-N\ln \alpha_{1} + O(1)
\end{equation}
which implies for the Gibbs free energy
\begin{equation}
G=-\frac{N}{\beta } \left( \frac{1}{2} \ln \frac{2m\pi }{\beta }
+\ln \frac{8}{\beta g^2 } -\ln \alpha_{1} \right)
\label{gfen}
\end{equation}
In view of (\ref{inire}) and (\ref{regf}),
$\alpha_{1} $ is the smallest solution of
\begin{equation}
\frac{\gamma +1}{z} = G^2 (z) =
\frac{1}{\displaystyle (\gamma +9)/z -1 -\frac{1}{\displaystyle
(\gamma +25)/z -1 -\frac{1}{\displaystyle (\gamma 
+ 49)/z -1 -\ldots } } }
\label{a1eq}
\end{equation}
For not too high pressures $P=g^2 (\gamma +1)/8 $, one can easily
obtain $\alpha_{1} $ numerically from this equation,
yielding the exact Gibbs free energy by inserting in (\ref{gfen}).
For high pressures, it is advantageous to use an asymptotic expansion
for $\alpha_{1} $, which will now be obtained along with an explicit low 
pressure expansion. By denoting $a=z-\gamma $ and $V_m = (a-m^2 )/z$,
(\ref{a1eq}) takes the form
\begin{equation}
0=V_1 -1 -\frac{1}{\displaystyle V_3 -\frac{1}{\displaystyle V_5 - 
\frac{1}{\displaystyle V_7 -\ldots } } }
\end{equation}
which is just the relation determining the characteristic values
corresponding to even periodic solutions of odd order
of the Mathieu equation, denoted in \cite{abr} as $a_{2r+1} (z) $.
Consequently, if $\gamma $ is given, then $\alpha_{1} $ is the
lowest solution of $a_{2r+1} (z) = z-\gamma $. The lowest solution
occurs for $r=0$, i.e. $\gamma = z-a_1 (z)$. Now, the limiting behavior
of $a_1 (z)$ is known and gives, translated back to $\alpha_{1} $ and $P$,
\begin{eqnarray}
\frac{8 P}{g^2 } &=& \frac{1}{8} \alpha_{1}^{2} +\frac{1}{64} \alpha_{1}^{3}
+\frac{1}{1536} \alpha_{1}^{4} -\frac{11}{36864} \alpha_{1}^{5}
-\frac{49}{589824} \alpha_{1}^{6}
+ \ldots \ \ \ \ \mbox{for} \ \alpha_{1} \rightarrow 0 \label{al0} \\
\frac{8 P}{g^2 } &=& 3 \alpha_{1} - 6\alpha_{1}^{1/2}
+\frac{9}{4} + \frac{3}{32} \alpha_{1}^{-1/2} + \frac{45}{256}
\alpha_{1}^{-1} + \ldots \ \ \ \ \mbox{for} \ \alpha_{1} \rightarrow
\infty \label{ali}
\end{eqnarray}
It is more intuitive to consider the equation of state, which can be
obtained by inverting (\ref{al0}) and (\ref{ali}), inserting in
\begin{equation}
V = \left. \frac{\partial G }{\partial P } \right|_{T} \ \ \ \ \  
\Rightarrow \ \ \ \ \ \rho T = \frac{NT}{V} =
\frac{\alpha_{1} }{(\partial \alpha_{1}
/ \partial P ) |_{T} }
\label{dicht}
\end{equation}
and again inverting the resulting expansions. In this way one arrives at
\begin{eqnarray}
P &=& \rho T \left( \frac{1}{2} -\frac{\sqrt{2} }{8}
\left( \frac{\rho T}{g^2 } \right)^{1/2} +
\frac{19}{96} \frac{\rho T}{g^2 } +\frac{73\sqrt{2} }{1536}
\left( \frac{\rho T}{g^2 } \right)^{3/2} -\frac{121}{1152}
\left( \frac{\rho T}{g^2 } \right)^{2} + \ldots \right) \label{lda} \\
P &=& \rho T \left( 1-\frac{\sqrt{6} }{4} \left( \frac{g^2 }{\rho T}
\right)^{1/2} + \frac{3}{32} \frac{g^2 }{\rho T} - \frac{15\sqrt{6} }{2048}
\left( \frac{g^2 }{\rho T} \right)^{3/2} +\frac{99}{8192}
\left( \frac{g^2 }{\rho T} \right)^{2} + \ldots \right) \label{hda}
\end{eqnarray}
The reason for the pressure only depending on the combination
$\rho T $ is discussed in \cite{prc}. The exact equation of state,
as obtained from (\ref{dicht}) in conjunction with (\ref{a1eq}),
is plotted in Fig. 1, compared with the case of particles in the
fundamental representation of SU(2) \cite{bam},\cite{me}.

Before discussing these results, it is instructive to repeat the above
calculation for the case that a particle in the fundamental representation
of SU(2) is positioned at either end of the system. This modifies the
possible set of color string configurations as follows: One starts
with one color string at one end of the array of adjoint particles 
and ends again with one color string at the other end, i.e.
$n_0^C = n_N^C =1$. One can never have less than one string, i.e.
$n_i^C \ge 1$. In all other ways, however, the combinatorics remains
the same; one has the same set of configurations except that the
number of strings is augmented by one everywhere. Therefore, one
arrives again at the recursion relation (\ref{recur}), 
and consequently also (\ref{regf}), except that
the index $r$ now runs over all odd numbers. Note also that at the
lowest level, $r=1$, one has no special constraint analogous to
(\ref{ct}) for the case $r=0$; one can couple one string, corresponding
to color spin one-half, with a color spin one adjoint particle back
to color spin one-half. Physically, in the color electric background
generated between a pair of separated fundamental sources, adjoint
particles are deconfined. By contrast, in the vacuum, adjoint
particles must combine to form color singlet clusters made up of two
or more particles.

The thermodynamic behavior of the system is now determined in complete
analogy to above by the pole $\beta_{1} $ of
$G^1 (z)$ nearest to the origin, i.e. the lowest solution of
\begin{equation}
\frac{\gamma +4}{z} -1 =
\frac{1}{\displaystyle (\gamma +16)/z -1 -\frac{1}{\displaystyle
(\gamma +36)/z -1 -\frac{1}{\displaystyle (\gamma
+ 64)/z -1 -\ldots } } }
\end{equation}
With the same abbreviations as above, this translates to the equation
\begin{equation}
0=V_2 -\frac{1}{\displaystyle V_4 -\frac{1}{\displaystyle V_6 -
\frac{1}{\displaystyle V_8 -\ldots } } }
\end{equation}
determining the characteristic values corresponding to odd periodic
solutions of even order of the Mathieu equation, denoted in \cite{abr}
as $b_{2r} (z)$. For given $\gamma $, one thus obtains $\beta_{1} $ as
the lowest solution of $b_{2r} (z) = z-\gamma $, which occurs for
$r=1$, i.e. $\gamma = z-b_2 (z)$. Using again the known limiting
behavior of $b_2 (z)$, one arrives at the following virial expansion
of the equation of state,
\begin{equation}
P = -\frac{3}{8} g^2 + \rho T \left( 1 -\frac{2}{3} \frac{\rho T}{g^2 }
+ \frac{16}{9} \left( \frac{\rho T}{g^2 } \right)^{2} -
\frac{145}{27} \left( \frac{\rho T}{g^2 } \right)^{3} +
\frac{1472}{81} \left( \frac{\rho T}{g^2 } \right)^{4} + \ldots \right) 
\label{ldb} 
\end{equation}
On the other hand, for large densities $\rho $, one has the same expansion
as in the case without fundamental sources (cf. eq. (\ref{hda})), since
the characteristic values of the Mathieu equation $b_2 $ and $a_1 $
have the same asymptotic expansion for large arguments $z$. Note
however that for any finite $z$, $a_1 $ and $b_2 $ differ \cite{abr}
and 
\begin{equation}
b_2 -a_1 \sim 2^9 \sqrt{2/\pi } z^{5/4} e^{-4\sqrt{z} } \ \ \ \ \ 
\mbox{for} \ z\rightarrow \infty
\label{guru}
\end{equation}
which of course does not show up in a $1/z$-expansion.

\section{Discussion}
The thermodynamic behavior displayed by the one-dimensional classical
Coulomb gas of particles in the adjoint representation of SU(2) can
be interpreted as follows: In the low density, low temperature limit,
the equation of state (\ref{lda}) yields the pressure corresponding to
a system of $N/2$ particles, since the $N$ adjoint constituents are
confined and pair off to form color singlet conglomerates in order
to minimize the amount of energy invested into color strings.
The behavior near this limit cannot be described by the usual virial
expansion in powers of the density $\rho $, as happens e.g. for
particles in the fundamental representation \cite{bam},\cite{me};
instead, one obtains an expansion in $\sqrt{\rho T/g^2 } $.
This is a signature for collective behavior under participation
of arbitrarily many particles; any cluster approximation taking into
account only interactions between a finite number of particles at a
time will necessarily lead to an expansion in powers of $\rho $.
The origin of this collective behavior is not difficult to pinpoint:
Adjoint particles have the possibility of forming multi-particle
clusters with only two color electric strings
extending over their entire length. These clusters are quite cheap
in terms of potential energy and provide a high level of degeneracy
already in the low-lying spectrum. Indeed, one can verify that by
taking into account only the aforementioned configurations, i.e.
approximating $G^2_n = 1/(\gamma +9)^n $, leading to
$G^2 (z) = z/(\gamma +9-z) $ in (\ref{a1eq}), one reproduces exactly
the first two terms in (\ref{lda}). Such an effect does not occur for
fundamental sources, since for these, string configurations of the type
described above are impossible. These configurations are analogous to
the states which have been
observed in \cite{kdb1}-\cite{zhi} to proliferate in the
spectrum of large-N$_{C} $ adjoint QCD$_{1+1} $, generating an
exponentially rising density of states leading to a Hagedorn
limiting temperature. One should realize that in the large-N$_{C} $
limit, the effect of these states is greatly enhanced by the fact that
the color strings are labeled by an additional infinity of quantum
numbers compared with the present case of SU(2). For SU(2), there is
no sharp phase transition; the crossover to the high density,
high temperature regime is smooth, as evidenced in Fig. 1.
This is to be expected on general grounds in a one-dimensional model;
only in the limit of infinitely many colors, where the infinity of
degrees of freedom at each space point acts like an extra dimension,
can one obtain a genuine deconfinement phase transition 
\cite{zar1}-\cite{zar3},\cite{kuta}. 

Nevertheless, in the present case of SU(2) color, the thermodynamic
behavior in the low density, low temperature limit is dominated by
the effect of arbitrarily large multi-particle clusters which only
contain two color strings along their entire length; as is to
be expected, the formation of these clusters initially leads to a
lowering of the pressure as compared with a gas of $N/2$ particles
composed of pairs of adjoint constituents. Only at higher densities
or temperatures does the equation of state turn around to approach
the high density, high temperature limit, cf. eq. (\ref{hda}), in
which the system behaves like an ideal gas of the $N$ adjoint
particles. In the presence of many nearby partners, an adjoint
particle may propagate on its own over large distances without
having to invest appreciable energy. The leading behavior away
from this deconfined limit is nonanalytic in the string tension,
namely proportional to $\sqrt{g^2 /(\rho T)} $. This signals the
presence of collective plasma excitations in the medium in analogy
to the Debye-H\"uckel law of electrodynamics. In this respect, the
gas of adjoint particles behaves no different from the gas of
particles in the fundamental representation \cite{bam},\cite{me}.

Turning now to the analysis of the case where a fundamental particle is
placed at either end of the system, one observes a negative pressure
$P=-3g^2 /8$ in the low density, low temperature limit, cf. eq. 
(\ref{ldb}). This is simply the force needed to keep the two fundamental
particles apart such that the system retains the desired
one-dimensional volume $V$ (note that in one dimension, a pressure
corresponds to a force since there is no transverse area to divide by).
The value $3g^2 /8$ is precisely the string tension associated with the
one color string extending between the fundamental sources. If one
did not exert this force, the system would strive to contract to
smaller volumes until the pressure of the adjoint particles equalized
the confining linear force between the two fundamental sources.
The leading behavior away from this limit corresponds to the pressure
generated by the $N$ adjoint deconfined particles. This stands in
accordance with the remark already made further above, that adjoint
particles are deconfined in the chromoelectric background generated
by the two fundamental sources, in contradistinction to the vacuum,
where they are confined and must form color singlet clusters.
For this same reason, also the collective effects stemming from the
formation of multi-particle clusters present in the vacuum are absent
in the chromoelectric background; the low density, low temperature
expansion of the equation of state takes the usual form containing
only powers of $\rho $, cf. eq. (\ref{ldb}).

On the other hand, in the high density, high temperature limit, the
pressure exerted by the system with two fundamental sources at the
ends is the same as the one exerted by the system without these sources,
to all orders in $\sqrt{g^2 /(\rho T)} $. Therefore, in this limit,
there is no vestige of the original confining force between the two
fundamental sources; the force is entirely screened and the sources
are deconfined. However, the expansion in $\sqrt{g^2 /(\rho T)} $ is
merely asymptotic; for any finite value of $\rho T/g^2 $, the pressure
of the system with fundamental sources is lower by an amount vanishing
exponentially in $\sqrt{\rho T/g^2 } $ for large $\rho T/g^2 $ in view
of eq. (\ref{guru}). In other words, the string tension between 
fundamental sources in the presence of the medium of adjoint particles
falls off exponentially for large $\rho T/g^2 $. In particular, there
is no entire regime in $\rho T/g^2 $ where the string tension vanishes
identically and which thus could be interpreted as a truly deconfined
phase.

\begin{figure}
\caption{Equation of state of the one-dimensional classical 
nonrelativistic Coulomb gas of particles in the adjoint
representation of SU(2) (solid line) and in the fundamental
representation of SU(2) (dashed).}
\end{figure}

\end{document}